\documentclass[11pt,preprint]{aastex}

\newcommand{\beq}{\begin{equation}}
\newcommand{\eeq}{\end{equation}}
\newcommand{\beqa}{\begin{eqnarray}}
\newcommand{\eeqa}{\end{eqnarray}}
\newcommand{\figref}[1]{Figure \ref{#1}}
\newcommand{\tabref}[1]{Table \ref{#1}}
\newcommand{\hmpc}{\,h^{-1} {\rm Mpc}}
\newcommand{\rmed}{r_{0\,\rm med}}
\newcommand{\npatch}{N_{\rm patch}}
\newcommand{\fivecols}{$UBVR_CI_C$}
\newcommand{\rc}{R_C}
\newcommand{\ic}{I_C}
\newcommand{\betam}{\beta_M}
\newcommand{\betaz}{\beta_z}
\newcommand{\mnot}{M_R^0}
\newcommand{\scut}{s_{\rm cut}}
\newcommand{\bias}{b_{\rm E}/b_{\rm L}}

\newcommand{\fig}[3]{%
\begin{figure}
\figurenum{#1}
\plotone{f#1.eps}
\caption{#3\label{#2}}
\end{figure}
\clearpage
}

\newcommand{\mmax}{21.5}
\newcommand{\rpmin}{0.04}
\newcommand{\rpmax}{10}
\newcommand{\rzmax}{25}
\newcommand{\brat}{1.7\pm0.2}
\newcommand{\bratsqapprox}{3}
\newcommand{\bratsq}{2.9\pm0.5}
\newcommand{\gamfid}{1.73}
\newcommand{\rscl}{r_0^{\gamma/1.73}}
\newcommand{\zminz}{0.12}
\newcommand{\zmaxz}{0.51}
\newcommand{\Mmaxz}{-20}
\newcommand{\Na}{1993}
\newcommand{\gama}{1.73\pm0.02}
\newcommand{\ra}{4.54\pm0.22}
\newcommand{\Ne}{720}
\newcommand{\game}{1.91\pm0.06}
\newcommand{\re}{5.45\pm0.28}
\newcommand{\Nl}{1273}
\newcommand{\gaml}{1.59\pm0.08}
\newcommand{\rl}{3.95\pm0.12}
\newcommand{\slopearz}{-2.1\pm1.3}
\newcommand{\slopeerz}{-3.9\pm1.0}
\newcommand{\slopelrz}{-7.7\pm1.3}
\newcommand{\zminm}{0.12}
\newcommand{\zmaxm}{0.40}
\newcommand{\Mmaxm}{-19.25}
\newcommand{\slopearm}{0.31\pm0.11}

\newcommand{\slopeerm}{0.35\pm0.17}
\newcommand{\negslopeerm}{-0.35\pm0.17}
\newcommand{\slopelrm}{-0.02\pm0.16}
\newcommand{\negslopelrm}{0.02\pm0.16}

\slugcomment{To appear in The Astrophysical Journal}
\shorttitle{The CNOC2 Galaxy Correlation Function}
\shortauthors{Shepherd et al.}

\begin{document}


\title{The Galaxy Correlation Function in the CNOC2 Redshift Survey:
Dependence on Color, Luminosity and Redshift}

\author{
C. W. Shepherd\altaffilmark{1},
R. G. Carlberg\altaffilmark{1,2},
H. K. C. Yee\altaffilmark{1,2},
S. L. Morris\altaffilmark{2,3}, 
H. Lin\altaffilmark{2,4,5},
M. Sawicki\altaffilmark{2,6},
P. B. Hall\altaffilmark{2,7},
and D. R. Patton\altaffilmark{1,2}
}

\altaffiltext{1}{Department of Astronomy and Astrophysics, University of
Toronto, Toronto, Ontario, M5S 3H8, Canada.  Email: carlberg, hyee, patton,
and shepherd@astro.utoronto.ca.}
\altaffiltext{2}{Visiting Astronomer, Canada-France-Hawaii Telescope, which is
operated by the National Research Council of Canada, Le Centre National de
Recherche Scientifique, and the University of Hawaii.}
\altaffiltext{3}{Department of Physics, University of Durham, Rochester
Building, Science Laboratories, South Road, Durham, DH1 3LE, England.  Email:
simon.morris@durham.ac.uk.}
\altaffiltext{4}{Steward Observatory, University of Arizona, Tucson, AZ, 85721.
Email: hlin@as.arizona.edu.}
\altaffiltext{5}{Hubble Fellow.}
\altaffiltext{6}{Department of Physics, California Institute of Technology,
Mail Code 320-47, Pasadena, CA, 91125.  Email:  sawicki@pirx.caltech.edu.}
\altaffiltext{7}{Princeton University Observatory and Pontificia Universidad
Cat\'{o}lica de Chile, Departamento de Astronom\'{\i}a y Astrof\'{\i}sica,
Facultad de F\'{\i}sica, Casilla 306, Santiago 22, Chile.
E-mail: phall@astro.puc.cl}


\begin{abstract}
%
%
We examine how the spatial correlation function of galaxies from the CNOC2
Field Galaxy Redshift Survey depends on galaxy color, luminosity and redshift.
The projected correlation function $w_p$ is determined for volume-limited
samples of objects with $\zminz\le z<\zmaxz$ and evolution-compensated
$\rc$-band absolute magnitudes $\mnot<\Mmaxz$, over the comoving projected
separation range $\rpmin\hmpc<r_p<\rpmax\hmpc$.
Our sample consists of 2937 galaxies which are classified as being either
early- or late-type objects according to their spectral energy distribution
(SED), determined from \fivecols\ photometry.
For simplicity, galaxy SEDs are classified independently of redshift: our
classification scheme therefore does not take into account the colour
evolution of galaxies.

%
%
Objects with SEDs corresponding to early-type galaxies are found to be more
strongly clustered by a factor of $\sim\bratsqapprox$, and to have a steeper
correlation function, than those with late-type SEDs.
Modeling the spatial correlation function, as a function of comoving
separation $r$, as $\xi(r)=\left(r/r_0\right)^{-\gamma}$, we find
$r_0=\re\hmpc$ and $\gamma=\game$ for early-type objects, and $r_0=\rl\hmpc$
and $\gamma=\gaml$ for late-type objects (for $\Omega_{\rm M}=0.2$,
$\Omega_\Lambda=0$).
While changing the cutoff between early- and late-type SEDs does affect the
correlation amplitudes of the two samples, the ratio of the amplitudes remains
constant to within 10\%.

%
%
The redshift dependence of the correlation function also depends on SED type.
Modeling the redshift dependence of the comoving correlation amplitude
$r_0^\gamma$ as $r_0^\gamma(z)\propto(1+z)^{\gamma-3-\epsilon}$, we find that
early-type objects have $\epsilon=\slopeerz$, and late-type objects have
$\epsilon=\slopelrz$.
Both classes of objects therefore have clustering amplitudes, measured in
comoving coordinates, which appear to decrease rapidly with cosmic time.
The excess clustering of galaxies with early-type SEDs, relative to late-type
objects, is present at all redshifts in our sample.
In contrast to the early- and late-type SED samples, the combined sample
undergoes little apparent evolution, with $\epsilon=\slopearz$, consistent
with earlier results.

%
%
The apparent increase with redshift of the clustering amplitude in the early-
and late-type samples is almost certainly caused by evolution of the
galaxies themselves, rather than by evolution of the correlation function.
If galaxy SEDs have evolved significantly since $z\sim0.5$, then our method of
classifying SEDs may cause us to overestimate the true evolution of the
clustering amplitude for the unevolved counterparts to our early- and
late-type samples.
However, if color evolution is to explain the apparent clustering evolution,
the color evolution experienced by a galaxy must be correlated with the galaxy
correlation function.

%
%
We also investigate the luminosity dependence of the correlation function for
volume-limited samples with $\zminm\le z<\zmaxm$ and $\mnot<\Mmaxm$.
We detect a weak luminosity dependence of the correlation amplitude for
galaxies with early-type SEDs, $d\log\xi/d\mnot=\negslopeerm$, but no
significant dependence for late-type objects, $d\log\xi/d\mnot=\negslopelrm$.

\end{abstract}



\section{Introduction}
\label{intro}

%
%
At low redshift, the clustering of galaxies is observed to depend both on
luminosity and on morphological type.
This behavior is expected if galaxies are biased tracers of the mass
distribution in the universe, since the amount of biasing for a particular
population depends on that population's mass and formation mechanism.
That clustering depends on galaxy morphology is generally accepted
\citep*[e.g.,][]{lov95,her96,wil98}.
Early-type galaxies are found to be more strongly clustered than late-type
objects; estimates of the ratio of correlation function amplitudes for the two
types range from $\sim1.5$ to $\sim5$.
The luminosity dependence of the correlation function is more contentious,
although results from most recent redshift surveys
\citep{lov95,her96,lin96,wil98,guz00} do indicate that high-luminosity
objects are more strongly clustered than low-luminosity galaxies, with
estimates of $d\log\xi/dM$ ranging from $0.1$ to $\gtrsim1$.

%
%
The luminosity and morphology dependence of the correlation function may vary
with redshift, however, as the evolution of bias may be different for objects
of different masses \citep{mo96,mat97,bau99}.
Luminosity- and type-dependent evolution is also seen in semi-analytic models
\citep{bau99,kau99}.
The emerging picture of slowly- or non-evolving galaxy clustering
\citep*{sma99,car00b,hog00} may therefore hide some important details.
Obtaining an empirical estimate of the evolution of the correlation function,
either as a function of luminosity, or of galaxy type, is extremely
challenging, however, due to the difficulty of defining a non-evolving
sample of galaxies.

%
%
We present here correlation function measurements for 2937 galaxies from the
Canadian Network for Observational Cosmology Field Galaxy Redshift Survey
\citep[CNOC2;][]{yee00}.
Attempts to measure the correlation function for different types of objects in
deep-redshift surveys have so far employed samples defined by a galaxy's
color as determined by a single color index \citep{lef96,car98} or on the
strength of the [\ion{O}{2}] line \citep{col94,sma99}.
Rather than rely on a single color index or line strength, we classify
galaxies according to their spectral energy distributions (SEDs), determined
from \fivecols\ photometry.
As opposed to most morphological classification systems, our SED-based
classification scheme has the additional advantage of being completely
quantitative.
The large number of objects in the survey allows us to investigate not only the
dependence of the correlation function on color, luminosity and redshift, but
also the joint color-luminosity and color-redshift dependences.

%
%
Our classification scheme does not account for the evolution of galaxy SEDs.
Therefore, both SED evolution and true evolution of the galaxy correlation
function will contribute to the apparent redshift dependence of the
correlation function in our SED-selected samples.
Despite this ambiguity, useful constraints on the evolution of the correlation
function may be obtained using this technique, independently of any particular
model for the evolution of galaxy SEDs.

%
%
This paper is organized as follows: In \S\ref{data}, we describe the data used
in our analysis.
Section \ref{corfunc} describes our technique for estimating the correlation
function from these data.
In \S\ref{results}, we describe how the correlation function depends on galaxy
color, redshift and luminosity; these results are discussed, and compared to
those from other works, in \S\ref{discussion}.
Throughout this paper, we take $\Omega_{\rm M}=0.2$, $\Omega_\Lambda=0$, and
$H_0=100h\,{\rm km\,s^{-1}Mpc^{-1}}$; absolute magnitudes are expressed using
$h=1$.
As the estimated correlation function depends on $\Omega_{\rm M}$ and
$\Omega_\Lambda$ through the angular separation-redshift relation, changing
these parameters would result in a different estimate of the evolution of the
correlation function.
However, \citet{car00b}, using a sample of data from the CNOC2 survey similar
to that used here, find that the correlation length changes by $\lesssim7\%$
when different cosmological parameters are assumed.
The variation of our results with cosmology is therefore likely to be small
compared to the statistical uncertainties.


\section{The CNOC2 Survey Data}
\label{data}

%
%
The CNOC2 Field Galaxy Redshift Survey is described in detail by
\citeauthor{yee00} (\citeyear{yee00}; also see \citealp*{yee96}).
We summarize here only the relevant aspects of the survey.

%
%
The sample consists of over 5000 redshifts for objects with $\rc<\mmax$.  The
rms velocity errors, determined from redundant observations, are
$\sim100{\rm\ km\, s^{-1}}$.
The data were obtained at the Canada-France-Hawaii Telescope (CFHT) using the
CFHT Multi-Object Spectrograph (MOS).
The survey covers four widely separated areas (``patches'') on the sky:
0223+00, 0920+37, 1447+09, 2148-05 (labeled by RA and Dec).
Each patch is roughly L-shaped, and comprises between 17 and 19 MOS fields,
each of which is approximately 9 arc-minutes by 8 arc-minutes in size.
The data used here are from the most recent version of the CNOC2 catalog,
a portion of which is available electronically: see \citet{yee00} for details.

%
%
As discussed in \S\ref{intro}, galaxy clustering is known to depend on
luminosity, with low-luminosity galaxies being less strongly correlated than
those with high luminosity.
This effect will cause the correlation function estimated from a
magnitude-limited sample to exhibit apparent redshift evolution, in addition to
any evolution caused by true changes of clustering, as only brighter objects
will be included in the sample at higher redshifts.
To avoid this effect, we consider only volume-limited samples in this analysis.

%
%
A further complication is that galaxies undergo substantial luminosity
evolution over the redshift range of the sample.
\citet{lin99} find that the characteristic magnitude, $M_\star$, of the
luminosity function for the CNOC2 sample evolves as $M_\star(z)=M_\star(0)-Qz$,
where the $\rc$-band evolution parameter $Q=1.51\pm0.53$ for objects with SED
number $s$ (defined in \S\ref{SED}) in the range $-0.5\le s<0.5$, $1.11\pm0.78$
for $0.5\le s<1.5$, and $0.22\pm0.76$ for $1.5\le s<3.5$.
Therefore, to sample the same population of objects at different redshifts, we
select galaxies based on their evolution-compensated $R_C$-band absolute
magnitudes $\mnot=M_R+Qz$.

%
%
The selection function for our volume-limited samples differs from unity, due
to varying observing conditions, non-uniform spatial sampling, MOS slit
collision, etc.
\citet{yee00} and \citet{yee96} have developed an accurate empirical estimate
of these observational selections as a function of galaxy magnitude, color,
position and redshift.
While we do employ these estimates in our calculation of the correlation
function, as discussed in \S\ref{corfunc}, we find that ignoring the selection
effects results only in small changes in the correlation function estimates,
typically less than 5\%.

\subsection{Spectral Energy Distribution Classification}
\label{SED}

%
%
In addition to the large number of redshifts in the survey, the CNOC2 survey
has the advantage of having \fivecols\ photometry for nearly all objects.
By comparing the observed colors to those predicted by redshifted template
spectra, we classify each object as having either an early- or a late-type
spectral energy distribution.
This allows us to investigate the evolution of the clustering of galaxies as a
function of SED class, rather than simply as a function of any one particular
color index or spectral feature.

%
%
The SED classification technique used here is the same as that used in
\citet{lin99} to determine the luminosity function for the CNOC2 survey.
We use template SEDs from \citeauthor*{col80} (\citeyear{col80}, hereafter
CWW).
We assign numerical values to the four CWW template SEDs as follows: template 0
is the Old Stellar Population (OSP) SED, taken to be the average of the M31 and
M81 OSP SEDs.
Templates 1, 2 and 3 are the CWW Sbc, Scd, and Im SED, respectively.
To determine the SED number $s$ of an object at a particular redshift, we
compute the apparent magnitudes for each of these four template SEDs, in each
of the five photometric bands, at the given redshift.
We then add $-0.05$ mag to $I_C$ for templates 0--2, in order to better match
the observed $\rc-\ic$ colors of the data.
We linearly interpolate the magnitudes between adjacent pairs of template SEDs,
and linearly extrapolate the (0,1) pair to negative numbers and the (2,3) pair
to numbers greater than 3.
These interpolated magnitudes are then fit to the observed magnitudes (after
correcting for Galactic reddening).
The result of the least-squares fit is the best-fitting SED number $s$ and
rest-frame absolute magnitudes in each of the photometric bands.

%
%
Having obtained an SED number $s$ for a galaxy, we classify it as having either
an early- or a late-type SED: early-type objects are those with $s<0.5$
(closest to the OSP SED); late-type objects have $s\ge0.5$ (closest to Sbc, Scd
or Im).
Out of the \Na\ objects in the survey which have $\mnot<\Mmaxz$ and $\zminz\le
z<\zmaxz$, \Ne\ are classified as having early-type SEDs, and \Nl\ as having
late-type SEDs.
Objects with $s<-0.5$ or $s\ge3.5$ have SEDs which do not closely match any of
the template SEDs and are therefore discarded; these objects constitute fewer
than 2\%
of the data.
Note that although our late-type sample corresponds to the combined
intermediate plus late-type sample in \citet{lin99}, we use the individual
intermediate- and late-type luminosity functions to compute the evolution
compensation $Qz$.

%
%
Our SED classification scheme does {\em not} account for evolution of galaxy
color.
An elliptical galaxy at redshift $\sim0.5$, for example, is expected to be
somewhat bluer than one at redshift zero.
If this color evolution is sufficiently strong, it is possible that a
high-redshift object will have an SED which is a closer match to a CWW template
SED of a later type than the true unevolved type of that object.
Strong color evolution would therefore cause our late-type sample to contain
progressively more E/S0 objects at higher redshifts, which could affect our
determination of the redshift-dependence of the correlation function.
This issue is discussed qualitatively in section \S\ref{discussion}, and will
be addressed more thoroughly in a future paper.


\section{Estimating the Correlation Function}
\label{corfunc}

%
%
We measure the spatial clustering of galaxies in the sample using the standard
projected correlation function $w_p(r_p)$, defined as
\beq
w_p(r_p)=\int_{-\infty}^\infty\xi(r_p,r_z)dr_z
\label{wpdef}
\eeq
\citep{peb80}.
Here, $\xi(r_p,r_z)$ is the two-point correlation function expressed as a
function of separation perpendicular to ($r_p$), and parallel to ($r_z$), the
line of sight to each pair of objects.
The parallel component of the observed separation of a pair of objects, $r_z$,
is the sum of their spatial separation and the relative peculiar velocity of
the pair, projected on to the line-of-sight to the pair.
As the integral in equation (\ref{wpdef}) is along the line-of-sight, $w_p$,
unlike $\xi(r_p,r_z)$, is an estimate of the spatial clustering of galaxies,
independent of their peculiar velocities.
If the spatial two-point correlation function is described by a power-law,
$\xi(r)=(r/r_0)^{-\gamma}$, then the projected correlation function will also
be a power-law:
\beq
w_p(r_p)=C_\gamma r_0 (r_p/r_0)^{1-\gamma}\;,
\label{wpmod}
\eeq
where $C_\gamma=\sqrt{\pi}\Gamma((\gamma-1)/2)/\Gamma(\gamma/2)$.

%
%
We estimate $\xi(r_p,r_z)$ as
\beq
\xi(r_p,r_z)={DD\cdot RR\over DR^2}-1\;,
\label{xiest}
\eeq
\citep{ham93}, where $DD$ is the weighted number of pairs of objects in the
sample with separation $(r_p,r_z)$.
The weights used for computing the pair counts are discussed in
\S\ref{weights}.
The $DR$ and $RR$ pair counts in equation (\ref{xiest}) are the weighted
number of data-random and random-random pairs, respectively, where the random
objects are drawn from a distribution which is unclustered, but otherwise
the same as that of the data.

%
%
The random catalog is constructed so as to have the same redshift, color and
magnitude distributions, and to be subject to the same selection effects, as
the data.
The luminosity function from \citet{lin99} determines the intrinsic joint
distribution in redshift, magnitude and SED class; the distribution of SED
numbers within an SED class is estimated directly from the data.
A preliminary, unselected random catalog is generated from this distribution.
Selection functions are then computed for each random object, by interpolating
the selection functions for the data in color and magnitude.
The final, selected random catalog is generated from the unselected catalog,
the selection function for an object determining the probability that that
object is included in the selected catalog.

%
%
Having obtained an estimate of $\xi(r_p,r_z)$ using equation (\ref{xiest}), we
integrate in the $r_z$ direction, as in equation (\ref{wpdef}).
We truncate the integral at $r_{z\,\max}=\rzmax\hmpc$; no significant variation
is seen in $w_p$ as $r_{z\,\max}$ is varied between $15\hmpc$ and $100\hmpc$.
The noise in $w_p$, however, increases substantially for $r_{z\,\max}
>\rzmax\hmpc$, as more physically unrelated pairs are added to the integral.

%
%
We estimate the uncertainty in the projected correlation function directly from
the data, using the patch-to-patch variation in $w_p$.
For a sample composed of $\npatch$ independent, identical patches, with
observed projected correlation functions $w_{p\,1},\dots,w_{p\,\npatch}$,
the variance in the mean,
\beq
\sigma^2_{w_p}={1\over\npatch(\npatch-1)}\sum_{i=1}^{\npatch}
    \left(w_{p\,i}-{1\over\npatch}\sum_{j=1}^{\npatch}w_{p\,j}\right)^2\;,
\label{wperr}
\eeq
is the best estimate of the square of the uncertainty in $w_p$, the projected
correlation function computed from all $\npatch$ patches combined.

%
%
Given estimates of $w_p$, and estimates of the uncertainty in $w_p$, the
correlation function parameters $r_0$ and $\gamma$ are determined from
a least-squares fit of equation (\ref{wpmod}) to the data.
The uncertainties in $r_0$ and $\gamma$ are also estimated from the
patch-to-patch variation, by fitting the data for each patch individually,
to obtain $r_{0\,i}$ and $\gamma_i$ for each patch $i$.
We then estimate the $1\sigma$ uncertainties in $r_0$ and $\gamma$ for the full
sample using equation (\ref{wperr}), with $w_p$ replaced by the appropriate
quantity.

\subsection{Pair Weights}
\label{weights}

%
%
The weight assigned to a pair of objects $(i,j)$ with separation $(r_p,r_z)$
in equation \ref{xiest} is $w_{ij} = w_iw_j$, where
\beq
w_i={1\over1+2\pi\bar n\phi_iJ_3(r_p,r_z)}\;.
\label{weight}
\eeq
Here, $\bar n$ is the mean number density of objects in the sample, $\phi_i$ is
the selection function for object $i$ and $2\pi J_3(r_p,r_z)$ is the integral
of the real-space correlation function over a cylinder of radius $r_p$ and
length $2r_z$.
When computed with these weights, the resulting pair count $DD$ will be the
number of statistically independent pairs of objects with given $r_p$ and $r_z$
separations, so that our weighting scheme is analogous to the minimum-variance
weighting suggested by \citet{ham93} and \citet{lov95} for the spatial
correlation function $\xi(r)$.

%
%
Since equation (\ref{weight}) depends on the correlation function we are trying
to estimate, via $J_3$, we must use an iterative technique to determine the
weights.
We first compute $DD(<r_p,<r_z)$ and $DR(<r_p,<r_z)$, the number of data-data
and data-random pairs, respectively, with separations $<r_p$ and $<r_z$, along
with $D$ and $R$, the number of data and random objects in the sample,
respectively.
These pair counts and object counts are computed with each object weighted by
the inverse of its selection function.
We then estimate the quantity $2\pi\bar nJ_3(r_p,r_z)$ as the average number of
objects in excess of random with separations $<r_p$ and $<r_z$ from a given
object:
\beq
\left(2\pi\bar nJ_3\right)_{\rm est}(r_p,r_z)
    ={DD(<r_p,<r_z)\over D}
    -{DR(<r_p,<r_z)\over R}\;,
\label{Jest}
\eeq
Finally, to estimate $\xi(r_p,r_z)$, we compute $DD$, $DR$, and $RR$ in
equation (\ref{xiest}), using the pair weights defined by equation
(\ref{weight}), with $2\pi\bar nJ_3$ estimated as above.

%
%
In principle, we could refine the pair weights, by using the estimated
correlation function and equation (\ref{wpmod}) to obtain an improved estimate
of $J_3$.
However, $J_3$ computed in this fashion differs from that estimated using
equation (\ref{Jest}) by less than $10\%$.
We consider this small inconsistency to be acceptable, as equation
(\ref{weight}) is only an approximation to the minimum-variance weighting
\citep{ham93}, and equation (\ref{xiest}) is an unbiased estimator of $\xi$,
regardless of the weights used to compute the pair counts.
Furthermore, there are no significant changes in the estimated correlations,
and only marginally larger estimated uncertainties, if $\xi$ is computed with
each object weighted by the inverse of its selection function, or assigned a
weight of unity.
Since the weights defined by equation (\ref{weight}) lie somewhere between
these two extreme cases, there is little point in further refinement of the
estimate of $J_3$.


\section{Results}
\label{results}

\subsection{Color Dependence}
\label{seddep}

%
%
To determine the color dependence of the correlation function, we consider
three samples: the early-type SED sample (consisting of objects with $-0.5\le
s<0.5$), the late-type sample ($0.5\le s<3.5$), and objects of all SED classes.
All three samples have $\zminz\le z<\zmaxz$ and $\mnot<\Mmaxz$.
These selection criteria yield volume-limited, evolution-compensated samples,
in the sense that any object in the sample would still have $\rc
=M_R+DM(z)+K(z)+Qz<\mmax$ if it were moved to any other redshift in the sample,
where $DM$ and $K$ are the usual distance modulus and K-correction,
respectively.
We compute $w_p$ as a function of projected comoving separation $r_p$ for each
SED sample, over the separation range $\rpmin\hmpc\le r_p<\rpmax\hmpc$.
Fitting equation (\ref{wpmod}) to the data, we obtain $r_0$ and $\gamma$ for
each of these samples.

%
%
Our results are shown in \figref{seddepfig}; the sample and fit parameters are
given in \tabref{seddeptab}.
The correlation functions for all three samples are well-fit by the power-law
model (\ref{wpmod}).
There is, however, some slow variation of the data about the model,
most evident in the late-type sample.
This is expected, as the measured values of $w_p$ at different separations are
not independent, since the same galaxies contribute to pairs at all
separations.
Thus, if the observed $w_p$ is greater (or less) than the universal value at
some particular separation $r_p$, then it is likely that the $w_p$ observed for
adjacent separation bins will also be greater (less) than the universal value. 
Random scatter in the data will therefore have the visual appearance of a slow,
non-random variation about the true correlation.

\placefigure{seddepfig}
\placetable{seddeptab}

%
%
Given our limited range of separations, therefore, it is difficult to assess
whether the model (\ref{wpmod}) is an adequate description of the data, or
whether some other model, in which $\gamma$ and $r_0$ vary with separation, is
appropriate.
However, even if the clustering is not well-described by a single power-law, 
the parameters $r_0$ and $\gamma$ from that model may be interpreted as the
correlation length and slope, averaged over the separation interval
$\rpmin\hmpc\le r_p<\rpmax\hmpc$.
In this case, the uncertainties determined using the method described in
\S\ref{corfunc} apply to these averaged quantities:  if $\gamma$ and $r_0$ are
thought to vary with separation, then their values at any particular
separation are not constrained by the error estimates for the
separation-averaged values.

%
%
\figref{seddepfig}\ clearly demonstrates the correlation function
depends strongly on SED class --- early-type objects are more strongly
clustered, and have a steeper slope, than late-type objects.
Therefore, at least one of these samples must be a strongly biased tracer of
the mass distribution.
We parameterize the relative bias between the the early- and late-type samples
according to the ratio of their correlation function amplitudes:
\beq
\bias=\sqrt{\left(r_0^\gamma\right)_{\rm E}/\left(r_0^\gamma\right)_{\rm L}}\;,
\label{biasdef}
\eeq
where the subscripts E and L denote quantities for the early- and late-type
samples, respectively.
The ratio of the early-type correlation function amplitude to that for the
late-type objects is $\bratsq$; the relative bias is therefore $\bias=\brat$.
Note that, since the early- and late-type correlation functions have different
slopes, the relative bias between the two samples is a power-law function of
separation; the quantity $\bias$, as defined above, measures the amplitude
of this function.

%
%
In \figref{scutfit}, we show how the amplitudes of the correlation functions
for the early- and late-type samples depend on the SED number, $\scut$, of the
cutoff between the early- and late-type SED classes.
We plot the quantity $\rscl$ vs. $\scut$, where $\rscl$ is the correlation
length scaled to $\gamma=\gamfid$, the value for the full sample from
\tabref{seddeptab}.
The scaled correlation length $\rscl$ is a measure of the amplitude of the
correlation function, and allows us to compare the correlation lengths of
different samples, independently of the slopes.
The uncertainties in $\rscl$ are computed in the same manner as those in $r_0$.

\placefigure{scutfit}

%
%
As expected, $\rscl$ for both samples decreases with $\scut$, as objects are
moved from the late-type sample to the early-type sample.
This effect, however, is small compared to the difference between the
clustering amplitudes of the two samples.
Moreover, the relative bias changes by $\lesssim10\%$
as $\scut$ is varied, and does not depend on $\scut$ in any systematic way.
Our results are therefore not sensitive to the choice of cutoff SED number, and
we adopt $\scut=0.5$ for the remainder of this paper.

\subsection{Redshift Dependence}
\label{zdep}

%
%
We next investigate how the correlation function for each SED class evolves
with redshift.
We divide each of the SED samples from \S\ref{seddep} into SED-redshift
subsamples, using the subsample parameters listed in \tabref{zdeptab}.
The redshift bins for each SED sample are chosen so that each bin contains
approximately the same number of objects.
This binning makes optimal use of the data; different redshift bins are
considered in \S\ref{missvar}.

\placetable{zdeptab}

%
%
In \figref{zssfig}, we plot the projected correlation function $w_p(r_p)$
for each of the SED-redshift subsamples defined in \tabref{zdeptab};
\figref{rzdepfig} summarizes the dependence of $\rscl$ on median redshift.
The excess correlation of the early-type objects relative to the late-type
objects is present at all redshifts.
Both the early- and late-type objects have a clustering amplitude which
{\em increases} with redshift.
This is somewhat surprising, since a scenario in which the correlation function
evolves due to gravitational instability would predict that the correlation
function should increase, not decrease, with cosmic time.
It therefore seems likely that the apparent evolution is actually evolution in
the bias of our SED-redshift subsamples relative to the matter distribution,
rather than an actual decay in the clustering of matter.
The cause of this evolution in bias is discussed in \S\ref{discussion}.

\placefigure{zssfig}
\placefigure{rzdepfig}

%
%
In contrast to the strong evolution of the early- and late-type samples, the
full sample undergoes little apparent evolution, consistent with earlier
results \citep{sma99,car00b,hog00}.
However, as can be seen in \tabref{zdeptab}, the ratio of the number of
early-type objects to the number of late-type objects in our sample varies
considerably with redshift; as similar variation is also present if the ratio
is computed by weighing each object by the inverse of its selection function.
The correlation function for data of all SED types will therefore exhibit
apparent evolution, arising from the variation in the composition of the sample
with redshift, in addition to any true evolution in clustering.
For our selection criteria, the combined sample is composed of approximately
equal numbers of early- and late-type objects at low redshifts, while at higher
redshifts it is dominated by late-type galaxies.
One would therefore expect that the correlation amplitude for the combined
sample would approximate the average of the amplitudes for the early- and
late-type samples at low redshifts, but more closely match the late-type
amplitude at higher redshifts, as is observed.

%
%
To quantify the apparent evolution of the correlation function, we begin with
the standard model
\beq
\xi_p(r;z)=\xi_p(r;0)(1+z)^{-(3+\epsilon)}\;,
\label{epsmodphys}
\eeq
\citep{gro77}, where $\xi_p(r;z)$ is the correlation function at physical, as
opposed to comoving, separation $r$, and redshift $z$.
In this model, $\epsilon=0$ corresponds to clustering fixed in physical
coordinates, and $\epsilon=\gamma-3$ corresponds to a constant comoving
correlation function.
We employ equations (\ref{wpmod}) and (\ref{epsmodphys}) to obtain a model for
the evolution of $\rscl$, the comoving correlation length scaled to our
fiducial slope, $\gamma=\gamfid$:
\beq
\rscl(z)=\rscl(0.3){(1+z)^{(\gamma(z)-3-\epsilon)/\gamfid}\over
                    1.3^{(\gamfid-3-\epsilon)/\gamfid}}\;,
\label{epsmod}
\eeq
where the quantity $\rscl(0.3)$ is the correlation length at $z=0.3$ scaled to
$\gamma=\gamfid$.
We use $z=0.3$, rather than $z=0$, as our fiducial redshift, since $\rscl(0)$
is poorly constrained by our data.

%
%
Equation (\ref{epsmodphys}) is a purely empirical model, and is not necessarily
expected to be valid at all redshifts.
In particular, this model predicts a monotonic evolution of $r_0$ with
redshift, at odds with both $N$-body simulation and observational
results \citep*{con98,bau99,kau99}.
Thus, while the model does adequately describe our data, its predictions
for $r_0$ at redshifts outside our observed range are to be treated with
suspicion.
Nonetheless, equation (\ref{epsmodphys}), applied to a limited range of
redshifts, is a potentially useful model, due to the physical significance of
the parameter $\epsilon$.

%
%
We fit equation (\ref{epsmod}) to the SED-redshift subsample data, taking $z$
to be the median redshift of the subsample.
The results are given in \tabref{zdepfittab}; error contours for the fits are
shown in \figref{rzdepfitfig}.
Both the early- and late-type data are best modeled by a correlation amplitude
which increases rapidly with redshift.
The late-type data exclude a constant clustering strength ($\epsilon=\gamma-3
=-1.27$) with $4\sigma$ confidence.
The apparent evolution in the early-type sample is less rapid, but still only
marginally consistent with $\epsilon=-1.27$.
The correlation amplitude for the full sample, on the other hand, undergoes
little apparent evolution, but is also consistent with $\epsilon=0$, or no
evolution in physical coordinates.

\placetable{zdepfittab}
\placefigure{rzdepfitfig}

%
%
Our interpretation of these results is based on the estimated uncertainties in
$\rscl$ and $\epsilon$ in \figref{rzdepfitfig}, which in turn depend on the
estimated uncertainties in $w_p$ from \tabref{zdeptab}.
The reliability of equation (\ref{wperr}), with only four independent patches,
is therefore an issue.
In particular, the small size of the error bar for the low-redshift points for
the early- and late-type SED-redshift subsamples, relative to the
higher-redshift uncertainties, is a cause for some concern:  insofar as the
SED-redshift subsamples are identical, one would expect that the uncertainties
in $\rscl$ would be the same for each subsample.
However, while the sizes of the subsamples are similar, the selection functions
are not: the higher-redshift subsamples are more sparsely sampled than their
lower-redshift counterparts, and therefore one would expect that they would
have larger uncertainties.
Furthermore, the uncertainty depends on the correlation function: strongly
clustered data will in general have a larger absolute uncertainty in the
correlation amplitude than will weakly clustered data.

%
%
We test how the small uncertainty in $\rscl$ at low redshift affects our
determination of $\rscl(0.3)$ and $\epsilon$ by re-fitting equation
(\ref{epsmod}) to the early- and late-type SED-redshift data, with the
uncertainties replaced by average values.
We compute the averaged uncertainties in two ways: first, as the
root-mean-square of the uncertainties in $\rscl$ from \tabref{zdeptab}, and
second, from the root-mean-square of the fractional uncertainties in $\rscl$.
In both cases, the best-fitting $\rscl$ and $\epsilon$ agree with the values
in \tabref{zdepfittab} to within a few percent, and the confidence regions are
actually somewhat smaller than those in \figref{rzdepfitfig}.
Our best uncertainty estimates in \tabref{zdeptab}, therefore, give the most
conservative uncertainty estimates for $\rscl$ and $\epsilon$.
We conclude that our interpretation of the results shown in
\figref{rzdepfitfig} is not unduly affected by  any unreliability in our
estimated uncertainties in $w_p$.

%
%
We next consider the apparent evolution of $\gamma$ with redshift.
In \figref{gamzdepfig}, we plot $\gamma$ vs. median redshift for each of the
SED-redshift samples.
We model the evolution using a simple linear model
\beq
\gamma(z)=\gamma(0.3)+\betaz(z-0.3)\;.
\label{gamzmod}
\eeq
The results of fitting equation (\ref{gamzmod}) to the data are given in
\tabref{zdepfittab}, and shown in \figref{gamzdepfitfig}.
We cannot place strong constraints on the evolution of $\gamma$ for either the
early- or the late-type SED samples: data from both samples are consistent with
constant $\gamma$, or with $\gamma$ being either a strongly increasing or a
decreasing function of redshift.
The full sample, on the other hand, does exhibit a significant decrease in
$\gamma$ with redshift.
As with the apparent redshift dependence of the correlation amplitude, however,
this may be due to the large difference between the slopes of the early- and
late-type correlation functions, and the redshift-dependent composition of the
combined sample, rather than true evolution of $\gamma$.

\placefigure{gamzdepfig}
\placefigure{gamzdepfitfig}

\subsection{Missing Variance}
\label{missvar}

%
%
By subdividing the sample into many subsamples, we run the risk of
underestimating the correlation function, due to the small volumes and numbers
of objects being considered \citep{ham93}.
We test the significance of this in several ways.
First, in \tabref{zdeptab}, we compare $r_0$ for each SED-redshift subsample to
the median, $\rmed$, of the 4 correlation lengths determined from each patch
individually, each scaled to the value of $\gamma$ for that subsample.
While $\rmed$ is typically somewhat smaller than $r_0$, in most cases this
difference is not significant at the $1\sigma$ level.
If the SED-redshift subsamples were small enough to be significantly affected
by missing variance, one would expect the individual patches to be affected
even more, and therefore have a significantly lower $\rmed$.

%
%
Second, we consider SED-redshift subsamples constructed using larger redshift
bins.
\figref{rzdepolapfig} shows $\rscl$ vs. redshift for the early- and late-type
SED classes, both using the redshift bins in \tabref{zdeptab}, and using
overlapping bins obtained by combining data from adjacent bins.
If the SED-redshift subsamples were significantly affected by missing variance,
one would expect the estimated correlation amplitude for the larger bins to be
greater than those for the two smaller bins.
Again, no significant difference in clustering amplitude is found: the value of
$\rscl$ for each overlapping bin is consistent with the average of the values
for the two bins it comprises.

\placefigure{rzdepolapfig}

%
%
Finally, we note that if cosmic variance in the correlation function is
significant for our subsamples, the fact that we use different redshift bins
for the three SED samples could affect our results, since the SED-redshift
subsamples sample different structures.
To test the significance of this effect, we compute $\rscl$ for the early-
and late-type samples with the same redshift bins as used for the combined
sample.
The results, plotted in \figref{rzdepsamefig}, are consistent with those
obtained using the preferred equal-number bins.
Since the clustering amplitude is not significantly affected by changing the
subsample sizes, or by sampling different structures, we conclude that missing
variance is not a serious problem in our sample.

\placefigure{rzdepsamefig}

\subsection{Luminosity Dependence}
\label{lumdep}

%
%
To determine the luminosity dependence of the correlation function, we consider
objects in the redshift range $\zminm\le z<\zmaxm$; the SED subsamples are
volume-limited for $\mnot<\Mmaxm$ over this redshift range.
We split these samples into $\mnot$ bins, again chosen so that each SED-$\mnot$
subsample has approximately the same number of objects.
The sample parameters are given in \tabref{Mdeptab}.

\placetable{Mdeptab}

%
%
\figref{Mssfig} shows $w_p(r_p)$ for each of the SED-$\mnot$ subsamples;
the dependence of $\rscl$ on median $\mnot$ is shown in \figref{rMdepfig}.
We model $\rscl(\mnot)$ as a power-law:
\beq
\rscl(\mnot)=\rscl(-20)\exp\left(-\kappa(\mnot+20)/\gamfid\right)\;.
\label{rmmod}
\eeq
In this model, $\rscl(-20)$ is the correlation length at $\mnot=-20$, for
$\gamma=\gamfid$.
The parameter $\kappa$ describes the fractional variation in the correlation
function amplitude with magnitude: $\kappa=-d\log\xi/d\mnot$.
The results of fitting  this model to the data, taking $\mnot$ to be the median
magnitude of the subsample, are listed in \tabref{Mdepfittab}; error contours
for these fits are shown in \figref{rMdepfitfig}.

\placefigure{Mssfig}
\placefigure{rMdepfig}
\placefigure{rMdepfitfig}
\placetable{Mdepfittab}

%
%
We find that the clustering amplitudes of the combined sample and of the
early-type sample are weakly increasing functions of luminosity: $\kappa
=\slopearm$ and $\slopeerm$, respectively.
The data from the late-type sample, on the other hand, are best modeled by an
amplitude which is essentially independent of luminosity ($\kappa
=\slopelrm$), although we cannot rule out a weak dependence on $\mnot$.
As with the redshift dependence of the correlation function, the apparent
luminosity dependence for the combined sample may be interpreted as being a
manifestation of the color dependence, due to the varying SED composition
of the sample with luminosity.

%
%
Finally, we consider variation of $\gamma$ with $\mnot$.
Figures \ref{gamMdepfig} and \ref{gamMdepfitfig} show $\gamma$ vs. median
$\mnot$ for the SED-$\mnot$ subsamples.
As for the redshift dependence of $\gamma$, we use a linear model for the
dependence on $\mnot$:
\beq
\gamma(\mnot)=\gamma(-20)+\betam(\mnot+20)\;.
\label{gammmod}
\eeq
The fit parameters $\gamma(-20)$ and $\betam$ are listed in
\tabref{Mdepfittab}.
As with the correlation amplitude, $\gamma$ for the early-type and combined
samples increases with luminosity at a moderate rate, while $\gamma$ for the
late-type sample is poorly constrained, but best modeled as being independent
of luminosity.
We note that, for the early-type and combined samples, $\gamma$ is
approximately independent of $\mnot$, except for the brightest subsamples, 
which have much larger values of $\gamma$ than the fainter subsamples.
Our simple model (\ref{gammmod}) may therefore not be entirely appropriate.

\placefigure{gamMdepfig}
\placefigure{gamMdepfitfig}


\section{Discussion}
\label{discussion}

\subsection{Color Dependence}
\label{seddepdis}

%
%
The strong dependence of clustering on SED indicates that galaxies are biased
tracers of the mass distribution at $z\sim0.4$, as is seen locally.
Results from the Stromlo-APM Redshift Survey \citep{lov95}, the Optical
Redshift Survey \citep{her96}, the Las Campanas Redshift Survey
\citep[LCRS;][]{lin96} and the Southern Sky Redshift Survey
\citep[SSRS2;][]{wil98} all indicate that, at $z\sim0$, early-type objects
are more strongly clustered, and have a steeper correlation function, than
late-type galaxies.
Quantitative comparisons between our results and those from these other surveys
are difficult, due to differing selection criteria, and differing methods of
quantifying the clustering.
Nonetheless, local estimates of the relative biasing $\bias$ range from 1.2 for
the SSRS2, to 2.3 for the Stromlo-APM survey, similar to our value of $\brat$.

%
%
A similar relative biasing has also been observed at higher redshifts.
In the Canada-France Redshift Survey (CFRS), \citet{lef96} find
$r_0 =2.1\pm0.2\hmpc$ (measured in physical coordinates) for galaxies redder
than the CWW Sbc SED, and $r_0=1.45\pm0.25\hmpc$ for blue objects over the
redshift range $0.2\le z\le0.5$, where both fits were performed with $\gamma$
fixed to 1.64.
These correlation lengths are considerably smaller than those found for our
data, possibly due to the small volumes and numbers of objects in the CFRS
survey.
The relative biasing of the two CFRS samples, however, is consistent with our
result.
Le F\`evre et al. find that the relative bias between their red and blue
samples is considerably smaller for $0.5\le z\le0.8$ than it is for
$0.2\le z\le0.5$.
As is clear from \figref{rzdepfig}, we detect no trend towards lower
relative biases at higher redshifts.

%
%
In the Norris survey, \citet{sma99} find $r_0=4.46\pm0.19\hmpc$ and
$\gamma=1.84\pm0.07$ for weak [\ion{O}{2}] emitters, and $r_0=3.80\pm0.40\hmpc$
and $\gamma=1.85\pm0.18$ for strong [\ion{O}{2}] sources, over the redshift
range $0.2\le z\le0.5$, for a relative biasing of $1.2$.
The Norris survey strong-[\ion{O}{2}] correlation amplitude is consistent with
our late-type amplitude, but the weak-[\ion{O}{2}] amplitude, and hence the
relative biasing, is considerably smaller than our value for the early-type
sample.
Assessing the significance of this discrepancy is difficult, however, due to
the different selection criteria.

\subsection{Luminosity Dependence}
\label{lumdepdis}

%
%
Quantitative comparisons between surveys of the luminosity dependence of the
correlation function are even more problematic than for the color dependence.
In addition to different selection criteria for the different surveys,
different authors determine the luminosity dependence of the correlation
function in different ways.
In the ESO Slice Project \citep[ESP;][]{guz00} and in the SSRS2, the luminosity
dependence of the correlation function is determined from a series of
concentric volume-limited samples of different sizes.
Results from both surveys indicate a strong luminosity dependence of the
correlation function amplitude: $\Delta\log\xi/\Delta M\sim-0.7$ for the
ESP, and $\Delta\log\xi/\Delta M\sim-1$ for the SSRS2, although there is
considerable variation with limiting magnitude for the latter survey.
These measurements are not directly comparable to our $\kappa$, since
$\Delta\log\xi/\Delta M$ expresses the fractional variation of the correlation
function with the magnitude limit of the sample.
One expects $\kappa$, determined from independent absolute magnitude bins, to
be larger than $\Delta\log\xi/\Delta M$ computed from overlapping samples, and
therefore the ESP and SSRS2 results indicate a luminosity dependence for the
correlation amplitude which is much stronger than we observe here.

%
%
Strong luminosity dependence is also seen in the LCRS and in the Norris Survey.
In these surveys, the luminosity dependence is determined by comparing the
clustering amplitude of objects brighter than some cutoff magnitude to that for
fainter objects.
In the LCRS, galaxies brighter than $M_\star-1$ are found to be $\sim50\%$
more strongly clustered than fainter objects.
For Norris Survey, the ratio of clustering amplitudes of objects brighter
than $M_\star$ to that for fainter objects is $\sim1.8$.
To compare these results to those from our survey, we average the clustering
amplitudes of our two brightest bins, and those from the two faintest bins.
The resulting cutoff magnitude, $\mnot=-20.10$, is close to $M_\star$ for
both early- and late-type galaxies.
We find that the bright galaxies are $\sim35\%$
more strongly clustered than fainter objects, a smaller difference than seen in
the LCRS and Norris surveys, but inconsistent with these results only at the
$1\sigma$ level.

%
%
The methodology for determining the luminosity dependence of the correlation
amplitude for the Stromlo-APM Survey most closely matches our own, as do the
results.
Using independent, volume-limited samples, selected jointly in absolute
magnitude and morphology, with $M_B<-17$, \citet{lov95} find
$\Delta\log\xi/\Delta M\sim-0.25$ for early-type objects, and
$\Delta\log\xi/\Delta M\sim-0.15$ for late type objects.
These values are reduced to $\sim-0.10$ for both morphological types, if the
sample is restricted to objects brighter than $M_B=-19.5$.
These measurements are in good agreement with our results, $d\log\xi/d M
=\negslopeerm$ and $\negslopelrm$ for objects with early- and late-type SEDs,
respectively.

\subsection{Redshift Dependence}
\label{zdepdis}

%
%
The unphysical evolution of the correlation function observed for the early-
and late-type SED samples is almost certainly caused by evolution of the
galaxies themselves, rather than true decay in clustering.
However, while galaxy evolution is found to cause the galaxy correlation
function to increase with redshift in semi-analytic models
\citep{bau99,kau99} this effect is generally limited to high redshifts.
For $z<0.5$, in their $\Lambda$CDM model (Figure 7), Kauffmann et al. find
that the correlation function of galaxies with high star formation rates
decreases rapidly with redshift, inconsistent with the evolution seen in our
late-type SED sample.
The correlation function of the Kauffmann et al. early-type morphology sample,
on the other hand, slowly increases with redshift at low redshifts.
While this behavior is qualitatively similar to that of our early-type SED
sample, the evolution in the Kauffmann et al. early-type morphology sample is
considerably less rapid than that in our sample.
As a result, the correlation amplitude of early-type galaxies predicted by
the model is $\sim50\%$ greater at $z=0.2$ than is observed here.
Of course, one would not necessarily expect the evolution of clustering in
SED-selected samples to match precisely that seen in samples selected by star
formation rate or by morphology.
The fact that the clustering in our SED-selected samples continues to decay at
low redshifts would therefore seem to indicate that the physical nature of
the galaxies in our samples is still evolving at $z\sim0.2$.

%
%
One form of galaxy evolution which can produce a correlation amplitude which
decreases with cosmic time is galaxy merging.
However, the merger rate varies approximately as $(1+z)^{-\epsilon}$, and so
would need to be much greater than is observed
(\citealp{car00a}; Patton et al. in preparation) in order to explain
our results.
Another, more likely, explanation is color evolution.

%
%
As discussed in \S\ref{SED}, the use of non-evolving template SEDs for galaxy
classification can result in some higher-redshift galaxies being systematically
misclassified as later types, so that our selection criteria do not select a
set of objects whose properties are independent of redshift.
Assuming our SED classifications correspond roughly to morphological types at
$z=0$, the early-type SED sample will comprise all E/S0 objects at low
redshifts, but at higher redshifts, only those ellipticals which do not undergo
significant color evolution.
The late-type SED sample, on the other hand, will contain objects of later
morphological types at all redshifts, but the high-redshift late-type
subsamples will also include those E/S0 objects which do undergo significant
color evolution.

%
%
A simple evolution scenario, in which the average color evolution experienced
by a galaxy depends only on that galaxy's unevolved SED, or its morphological
type, probably cannot explain the apparent evolution of the correlation
function, however.
Misclassifying some high-redshift elliptical galaxies as having later-type
SEDs would cause an apparent increase with redshift in the correlation
amplitude for our late-type SED sample, but would not affect the observed
evolution of the early-type clustering.
Although it is unlikely, since the CWW template SEDs are compiled from the
spectra of local galaxies, the template SEDs could also be too blue, in which
case our low-redshift early-SED subsamples would contain some galaxies of later
unevolved types.
While this would explain the apparent decrease with cosmic time of the
early-type correlation amplitude, it would also result in a relative bias
between unevolved early- and late-type objects at low redshifts which is much
greater than is observed.

%
%
If color evolution is to explain the apparent clustering evolution, we require
that the color evolution experienced by a galaxy be correlated with the
galaxy correlation function.
As an example, suppose that E/S0 objects which undergo sufficient color
evolution to be classified as having late-type SEDs at $z\sim0.5$ are for some
reason less strongly clustered than those ellipticals with non-evolving colors
(but still more strongly clustered than objects of later morphological types).
Both the early- and late-type high-redshift SED subsamples will therefore
contain objects which are more strongly clustered than those in their
low-redshift counterparts, resulting in an apparent increase with redshift of
the correlation function amplitude for both types.

%
%
One possibility is that E/S0 objects which undergo significant color evolution
have smaller masses than those whose SEDs do not evolve.
Since low-mass objects are expected to cluster less strongly than high mass
objects, E/S0 galaxies with non-evolving SEDs would be more strongly clustered
than those with evolving SEDs, resulting in the observed evolution of
$r_0^\gamma$.
Assuming that more massive E/S0 objects are also more luminous, this scenario
is consistent with the apparent luminosity dependence of the correlation
function, and could also explain the evolution of the luminosity function, as
reported in \citet{lin99}, where the early-type SED sample undergoes strong
luminosity evolution, while the late-type sample is subject to strong number
evolution.

%
%
Another possible connection between color evolution and clustering is galaxy 
age.
Elliptical galaxies which formed sufficiently early in the history of the
universe have presumably stopped evolving by $z\sim0.5$.
E/S0 objects formed at later epochs, on the other hand, might still have
significantly bluer SEDs at $z\sim0.5$ than those at $z=0$.
However, galaxies which form at early epochs are expected to be more strongly
clustered than those formed at later epochs, since the peaks in the matter
distribution are rarer at higher redshifts than at lower redshifts, and
therefore more strongly biased \citep{kai84}.
Older ellipticals experiencing less color evolution than young ones could
therefore also explain the observed evolution of $r_0^\gamma$.

%
%
Of course, other mechanisms may be responsible for the relationship between the
correlation function and the color evolution rate, and the relationship may
not be that described in the example above.
We stress, however, that regardless of the details, if color evolution is the
cause of the apparent evolution of the correlation function, then the color
evolution rate must be in some way correlated with the galaxy correlation
function.
This argument may be inverted:  if the apparent evolution of clustering of a
sample of galaxies is to be assigned a clear physical interpretation, then the
selection criteria of that sample must be based on galaxy properties which are
either independent of redshift, or depend on redshift in a manner which is
unrelated to the correlation function.
The foregoing considerations make clear that defining such a sample is likely
to be difficult, since the galaxy properties used to define the sample (such
as luminosity and SED) are quite likely to evolve, and it is plausible that
the evolution of these properties does depend on the correlation function.


\section{Summary}
\label{summary}

%
%
We have investigated how the correlation function depends on galaxy color,
redshift and luminosity, using volume-limited, evolution-compensated samples
from the CNOC2 redshift survey.
Galaxies are classified according to their spectral energy distributions, using
non-evolving template SEDs.
The samples considered appear to be large enough to avoid being significantly
affected by missing variance.

%
%
The correlation function depends strongly on SED: early-type objects are more
strongly clustered, and have a steeper correlation function slope, than
late-type objects.
We find $r_0=\re\hmpc$, $\gamma=\game$ for early-type objects, and $r_0
=\rl\hmpc$, $\gamma=\gaml$ for late-type galaxies, over the redshift range
$\zminz\le z<\zmaxz$.
The combined sample (both early- and late-type objects) has $r_0=\ra\hmpc$
and $\gamma=\gama$.
The relative bias of the early- and late-type samples is $\bias=\brat$; a large
relative bias exists at all redshifts in the sample.

%
%
The correlation amplitude for early-type objects increases at a moderate rate
with luminosity, $d\log\xi/d\mnot=\negslopeerm$, while that for late-type
galaxies has no significant luminosity dependence:
$d\log\xi/d\mnot=\negslopelrm$.
The slope of the early-type correlation function also depends on luminosity,
with $\gamma$ being largest for the brightest subsample; $\gamma$ for the
late-type sample is best modeled as being independent of luminosity, although
this result is poorly constrained.
The luminosity dependence of the correlation function for the combined sample
is similar to that for the early-type sample.
However, the SED composition of the combined sample varies with luminosity, and
so the apparent luminosity dependence of this sample is at least in part caused
by the strong color dependence.

%
%
Both early- and late-type objects have clustering amplitudes which appear to
increase rapidly with redshift: $\epsilon=\slopeerz$ for early-type objects,
$\epsilon=\slopelrz$ for late types.
While the dependence of $\gamma$ on $z$ for both these samples is consistent
with no evolution, we cannot place strong constraints on its behavior.
The combined sample, on the other hand, shows little evolution in amplitude,
with $\epsilon=\slopearz$, but a significant decrease in $\gamma$ with
redshift.
This apparent discrepancy may be explained by the varying SED composition of
the full sample with redshift.

%
%
If our SED-redshift subsamples represent populations of objects with physical
properties which are independent of redshift, the observed evolution of the
amplitude of the correlation function is inconsistent with a scenario in which
gravitational instability drives the formation and evolution of structure.
However, if galaxy colors have undergone significant evolution since $z
\sim0.5$, then our use of non-evolving template SEDs may mean that our
selection criteria are not in fact independent of redshift.
If the color evolution experienced by a galaxy is correlated with the galaxy
correlation function, then color evolution could explain the apparent increase
in clustering amplitude with redshift.

%
%
Even if colour evolution is significant, our results on the joint SED-redshift
dependence of the correlation function, as shown in \figref{rzdepfig}, still
serve as a model-independent description of the apparent evolution of galaxy
clustering as a function of galaxy SED.
Two important conclusions may be drawn from these results, independent of any
particular colour evolution model.
Firstly, there is a large range of clustering bias over the entire redshift
range $\zminz\le z<\zmaxz$.
Secondly, galaxy evolution makes a major contribution to the apparent evolution
of galaxy clustering.
It is therefore extremely difficult to construct a galaxy sample which
traces the mass distribution in a way which is independent of redshift, as is
necessary if physical conclusions about the evolution of the mass correlation
function are to be obtained directly from galaxy survey data.


\acknowledgements

This project was supported by a collaborative program grant from NSERC, as well
as individual operating grants from NSERC to RGC and HKCY.  HL Acknowledges
support provided by NASA through Hubble Fellowship grant \#HF-01110.01-98A
awarded by the Space Telescope Science Institute, which is operated by the
Association of Universities for Research in Astronomy, Inc., for NASA under
contract NAS 5-26555.


\break


\begin{deluxetable}{lcrcc}
\tabletypesize{\small}
\tablewidth{0pt}
\tablecaption{
SED Sample Correlation Function Parameters
\label{seddeptab}}
\tablehead{
\colhead{SED\tablenotemark{a}} &
\colhead{$z_{\rm med}$} &
\colhead{$N$} &
\colhead{$r_0$\tablenotemark{b}} &
\colhead{$\gamma$\tablenotemark{b}}}
\startdata
  All & 0.375 & 1993 & $4.54\pm0.22$ & $1.73\pm0.02$\\
Early & 0.338 &  720 & $5.45\pm0.28$ & $1.91\pm0.06$\\
 Late & 0.392 & 1273 & $3.95\pm0.12$ & $1.59\pm0.08$\\
\enddata
\tablenotetext{a}{All samples have $\mnot<-20$ and $0.12\le z<0.51$.}
\tablenotetext{b}{One-parameter $1\sigma$ confidence intervals.}
\end{deluxetable}

\begin{deluxetable}{llcccccc}
\tabletypesize{\small}
\tablewidth{0pt}
\tablecaption{
SED-Redshift Subsample Correlation Function Parameters
\label{zdeptab}}
\tablehead{
\colhead{SED\tablenotemark{a}} &
\colhead{$z_{\rm min},z_{\rm max}$} &
\colhead{$z_{\rm med}$} &
\colhead{$N$\tablenotemark{b}} &
\colhead{$r_0$\tablenotemark{c}} &
\colhead{$\gamma$\tablenotemark{c}} &
\colhead{$\rscl$\tablenotemark{c}} &
\colhead{$\rmed$\tablenotemark{d}}}
\startdata
All
& 0.120, 0.292 & 0.228 & 499 (275+224)
    & $3.75\pm0.23$ & $1.87\pm0.15$ & $4.17\pm0.49$ & 3.73\\
& 0.292, 0.375 & 0.350 & 493 (189+304)
    & $3.87\pm0.64$ & $1.82\pm0.16$ & $4.15\pm0.47$ & 3.28\\
& 0.375, 0.438 & 0.398 & 502 (156+346)
    & $4.26\pm0.15$ & $1.65\pm0.07$ & $3.98\pm0.28$ & 4.05\\
& 0.438, 0.510 & 0.469 & 499 (100+399)
    & $5.14\pm0.56$ & $1.53\pm0.09$ & $4.25\pm0.29$ & 4.96\\
Early
& 0.120, 0.270 & 0.221 & 248
    & $4.12\pm0.20$ & $2.05\pm0.08$ & $5.35\pm0.20$ & 3.90\\
& 0.270, 0.382 & 0.350 & 234
    & $5.30\pm1.04$ & $1.95\pm0.16$ & $6.55\pm1.16$ & 4.28\\
& 0.382, 0.510 & 0.426 & 238
    & $4.96\pm0.62$ & $2.10\pm0.11$ & $6.99\pm0.57$ & 4.63\\
Late
& 0.120, 0.324 & 0.258 & 321
    & $2.41\pm0.24$ & $1.68\pm0.08$ & $2.35\pm0.23$ & 2.61\\
& 0.324, 0.392 & 0.366 & 314
    & $3.24\pm0.60$ & $1.74\pm0.30$ & $3.26\pm0.82$ & 3.05\\
& 0.392, 0.453 & 0.418 & 322
    & $4.51\pm0.70$ & $1.60\pm0.08$ & $4.03\pm0.71$ & 3.82\\
& 0.453, 0.510 & 0.473 & 316
    & $4.65\pm0.53$ & $1.59\pm0.12$ & $4.11\pm0.25$ & 4.29\\
\enddata
\tablenotetext{a}{All subsamples have $\mnot<-20$.}
\tablenotetext{b}{The numbers in parentheses for the combined subsamples are
the number of early-type and late-type galaxies comprised by the
subsample.}
\tablenotetext{c}{One-parameter $1\sigma$ confidence intervals.}
\tablenotetext{d}{The median of the scaled correlation lengths from the
individual patches. See \S\ref{missvar} for details.}
\end{deluxetable}

\begin{deluxetable}{lcrcr}
\tabletypesize{\small}
\tablewidth{0pt}
\tablecaption{
Correlation Function Redshift Dependence
\label{zdepfittab}}
\tablehead{
\colhead{SED} &
\colhead{$\rscl(z=0.3)$\tablenotemark{a}} &
\colhead{$\epsilon$\tablenotemark{a}} &
\colhead{$\gamma(z=0.3)$\tablenotemark{a}} &
\colhead{$\betaz$\tablenotemark{a}}}
\startdata
  All & $4.1\pm0.3$ & $-2.1\pm1.3$ & $1.8\pm0.1$ & $-1.5\pm0.7$\\
Early & $5.7\pm0.2$ & $-3.9\pm1.0$ & $2.0\pm0.1$ & $ 0.1\pm0.6$\\
 Late & $2.7\pm0.2$ & $-7.7\pm1.3$ & $1.7\pm0.1$ & $-0.4\pm0.6$\\
\enddata
\tablenotetext{a}{The tabulated uncertainties are the one-parameter $1\sigma$
confidence intervals.  See Figures \ref{rzdepfitfig} and \ref{gamzdepfitfig}
for the joint two-parameter confidence regions.}
\end{deluxetable}

\begin{deluxetable}{llccccc}
\tabletypesize{\small}
\tablewidth{0pt}
\tablecaption{
SED-$\mnot$ Subsample Correlation Function Parameters
\label{Mdeptab}}
\tablehead{
\colhead{SED\tablenotemark{a}} &
\colhead{$M^0_{R\,\rm min},M^0_{R\,\rm max}$\tablenotemark{b}} &
\colhead{$M^0_{R\,\rm med}$} &
\colhead{$N$\tablenotemark{c}} &
\colhead{$r_0$\tablenotemark{d}} &
\colhead{$\gamma$\tablenotemark{d}} &
\colhead{$\rscl$\tablenotemark{d}}}
\startdata
All
& -22.52, -20.60 & -20.99 & 576 (292+284)
    & $3.68\pm0.31$ & $1.92\pm0.06$ & $4.25\pm0.28$\\
& -20.60, -20.10 & -20.31 & 554 (215+339)
    & $3.74\pm0.25$ & $1.68\pm0.09$ & $3.60\pm0.23$\\
& -20.10, -19.65 & -19.88 & 557 (188+369)
    & $3.25\pm0.27$ & $1.81\pm0.13$ & $3.43\pm0.16$\\
& -19.65, -19.25 & -19.46 & 542 (113+429)
    & $3.52\pm0.51$ & $1.65\pm0.08$ & $3.32\pm0.44$\\
Early
& -22.52, -20.65 & -21.11 & 278
    & $4.66\pm0.34$ & $2.14\pm0.02$ & $6.71\pm0.53$\\
& -20.65, -20.00 & -20.27 & 276
    & $4.68\pm0.08$ & $1.89\pm0.06$ & $5.40\pm0.19$\\
& -20.00, -19.25 & -19.68 & 254
    & $5.19\pm0.73$ & $1.85\pm0.08$ & $5.82\pm0.81$\\
Late
& -22.27, -20.45 & -20.83 & 352
    & $3.11\pm0.36$ & $1.58\pm0.25$ & $2.82\pm0.27$\\
& -20.45, -20.00 & -20.21 & 359
    & $2.86\pm0.18$ & $1.71\pm0.16$ & $2.83\pm0.26$\\
& -20.00, -19.60 & -19.79 & 345
    & $3.22\pm0.55$ & $1.67\pm0.16$ & $3.09\pm0.54$\\
& -19.60, -19.25 & -19.42 & 365
    & $2.95\pm0.27$ & $1.67\pm0.15$ & $2.84\pm0.27$\\
\enddata
\tablenotetext{a}{All subsamples have $0.12\le z<0.40$.}
\tablenotetext{b}{$M^0_{R\,\rm min}$ for the brightest bin for each SED class
is taken to be the $\mnot$ of the brightest object of that class.}
\tablenotetext{c}{The numbers in parentheses for the combined subsamples are
the number of early-type and late-type galaxies comprised by the
subsample.}
\tablenotetext{d}{One-parameter $1\sigma$ confidence intervals.}
\end{deluxetable}

\begin{deluxetable}{lcrcr}
\tabletypesize{\small}
\tablewidth{0pt}
\tablecaption{
Correlation Function Absolute Magnitude Dependence
\label{Mdepfittab}}
\tablehead{
\colhead{SED} &
\colhead{$\rscl(\mnot=-20)$\tablenotemark{a}} &
\colhead{$\kappa$\tablenotemark{a}} &
\colhead{$\gamma(\mnot=-20)$\tablenotemark{a}} &
\colhead{$\betam$\tablenotemark{a}}}
\startdata
  All & $3.50\pm0.12$ & $ 0.31\pm0.11$ & $1.73\pm0.04$ & $-0.17\pm0.06$\\
Early & $5.19\pm0.24$ & $ 0.35\pm0.17$ & $1.87\pm0.04$ & $-0.24\pm0.05$\\
 Late & $2.85\pm0.14$ & $-0.02\pm0.16$ & $1.67\pm0.08$ & $ 0.03\pm0.18$\\
\enddata
\tablenotetext{a}{The tabulated uncertainties are the one-parameter $1\sigma$
confidence intervals.  See Figures \ref{rMdepfitfig} and \ref{gamMdepfitfig}
for the joint two-parameter confidence regions.}
\end{deluxetable}


\newpage

\fig{1}{seddepfig}
{Projected comoving correlation function $w_p(r_p)$ for objects with $\mnot
<\Mmaxz$ and $\zminz\le z<\zmaxz$, from all SED classes, and the early- and
late-type SED classes individually.  The lines show the model (equation
\ref{wpmod}), fit to the data for each sample.  The fit parameters are given
in \tabref{seddeptab}.}

\fig{2}{scutfit}
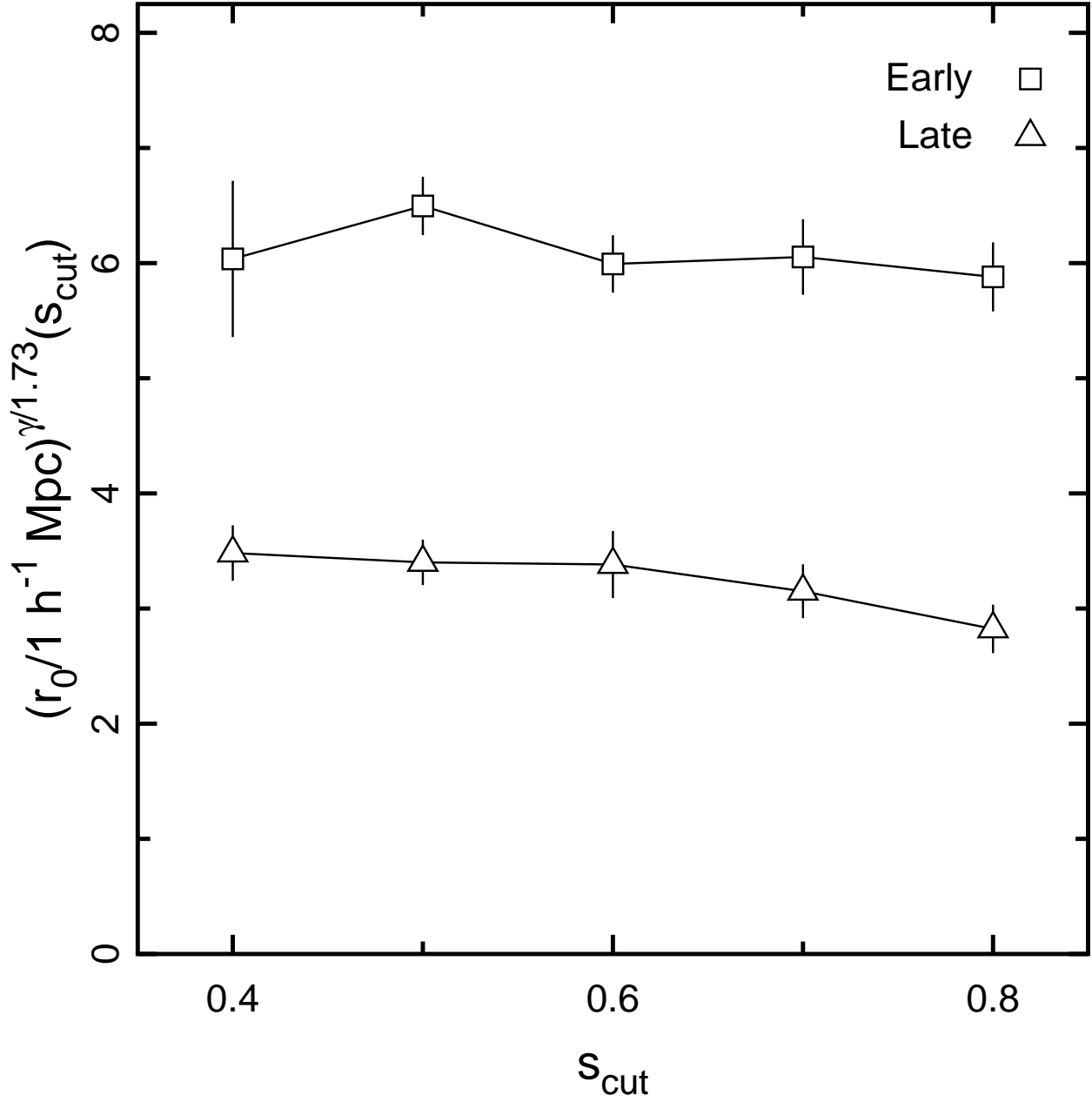
{Scaled comoving correlation length $\rscl$ vs. cutoff SED number $\scut$ for
the early- and late-type SED samples.  For each cutoff $\scut$, the early- and
late-type SED samples consist of those objects with $-0.5\le s<\scut$, and
$\scut\le s<3.5$, respectively.}

\fig{3}{zssfig}
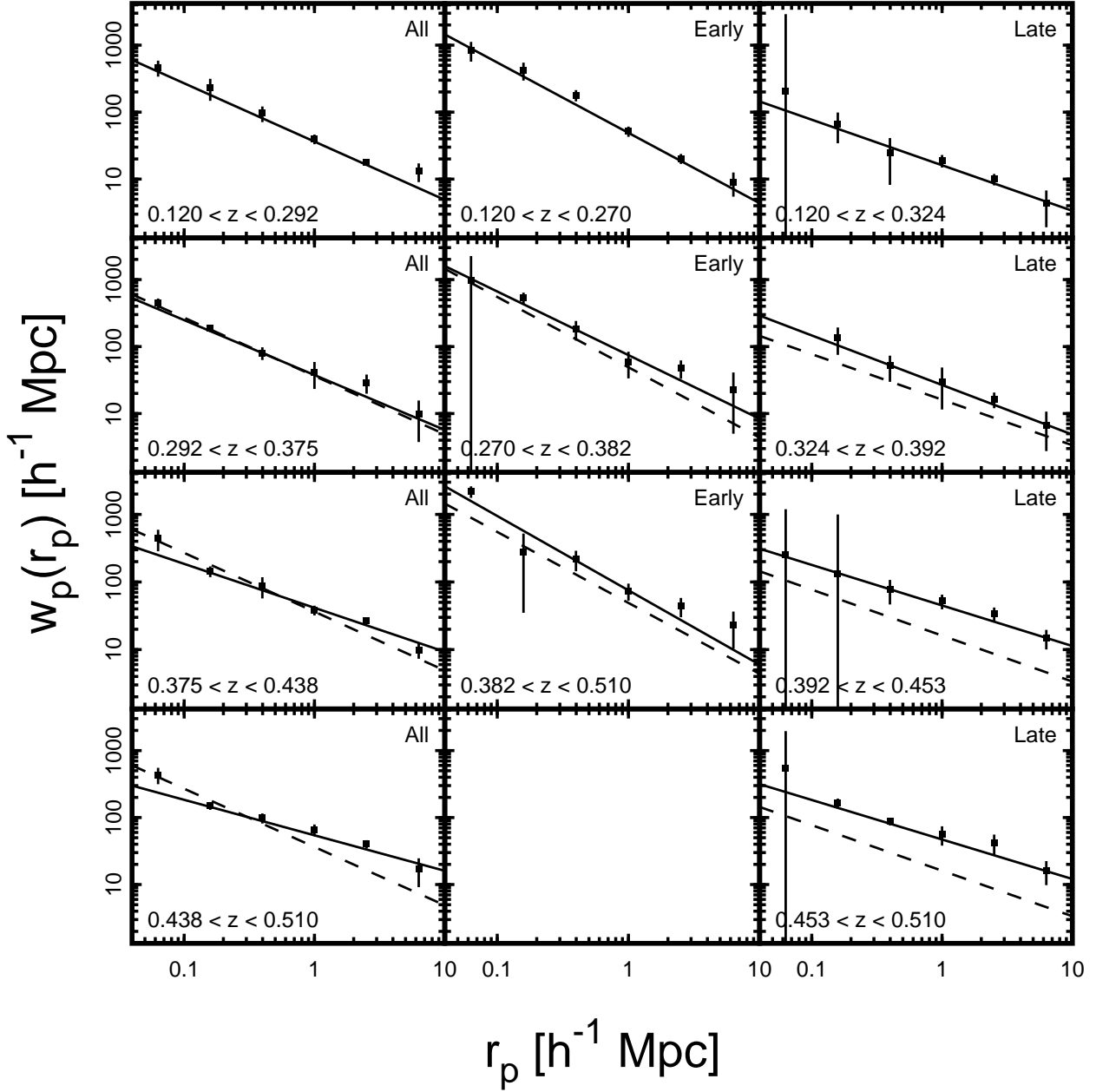
{Projected comoving correlation function $w_p(r_p)$ for each of the
SED-redshift subsamples.  The solid line in each panel is the model (equation
\ref{wpmod}), fit to the data for that subsample; the fit parameters are given
in \tabref{zdeptab}.  The dashed lines show the model from the lowest-redshift
subsample for each SED class.}

\fig{4}{rzdepfig}
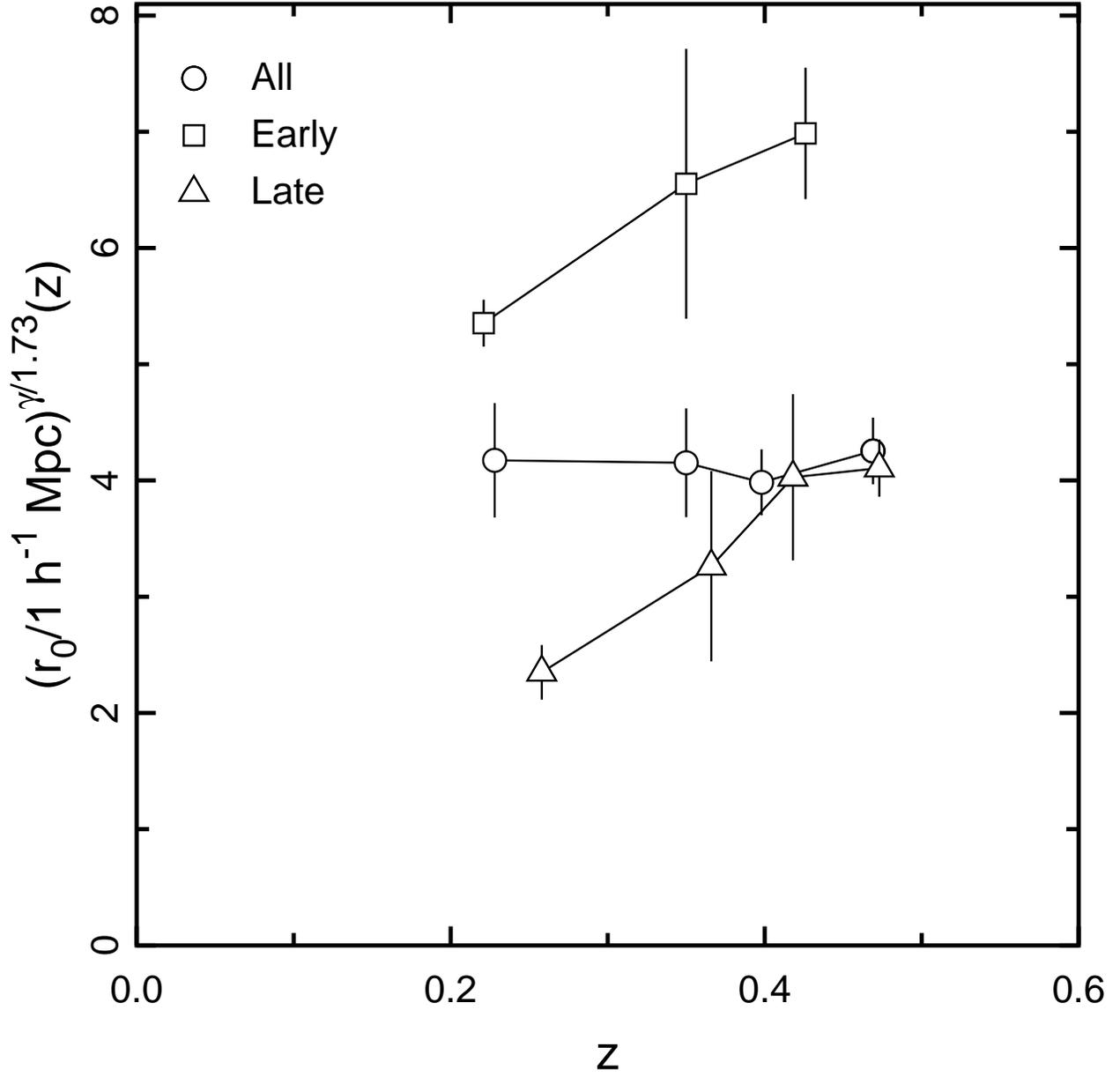
{Scaled comoving correlation length $\rscl$ vs. median redshift for each
SED-redshift subsample.  Each subsample has $\mnot<\Mmaxz$.}

\fig{5}{rzdepfitfig}
{One- and $2\sigma$ ($\Delta\chi^2=$ 2.30 and 6.17) error contours for the fits
of equation (\ref{epsmod}) for the SED-redshift subsamples.  The leftmost
vertical line corresponds to $\epsilon=\gamfid-3=-1.27$, or no evolution in
comoving coordinates.  The rightmost vertical line is $\epsilon=0$, or no
evolution in physical coordinates.}

\fig{6}{gamzdepfig}
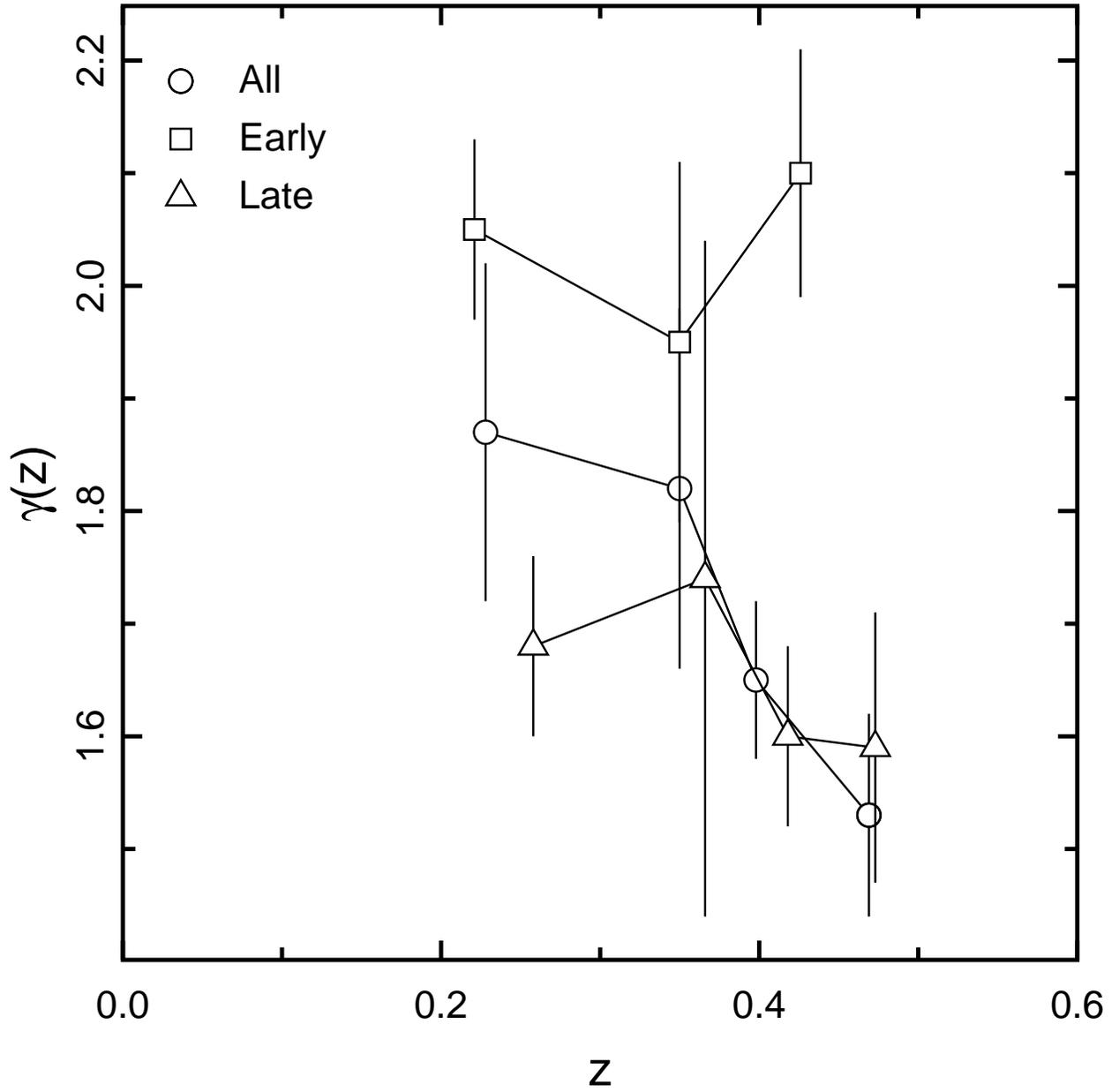
{Correlation function slope $\gamma$ vs. median redshift for the SED-redshift
subsamples.}

\fig{7}{gamzdepfitfig}
{One- and $2\sigma$ ($\Delta\chi^2=$ 2.30 and 6.17) error contours for the fits
of equation (\ref{gamzmod}) for the SED-redshift subsamples.}

\fig{8}{rzdepolapfig}
{Scaled comoving correlation length $\rscl$ for the early- and late-type
SED-redshift subsamples, for different choices of redshift bins.  The open
symbols are $\rscl$ computed using the redshift bins listed in
\tabref{zdeptab}; these are the same data as shown in \figref{rzdepfig}.  The
filled symbols are $\rscl$ computed for overlapping redshift bins constructed
by combining data from adjacent bins --- see \S\ref{missvar} for details.}

\fig{9}{rzdepsamefig}
{Scaled comoving correlation length $\rscl$ for the early- and late-type
SED-redshift subsamples, for different choices of redshift bins.  The open
symbols are $\rscl$ computed using the redshift bins listed in
\tabref{zdeptab} for the appropriate SED class; these are the same data as
shown in \figref{rzdepfig}.  The filled symbols are $\rscl$ for the early- and
late-type samples, computed using the redshift bins for the combined sample.}

\fig{10}{Mssfig}
{Projected comoving correlation function $w_p(r_p)$ for each of the
SED-$\mnot$ subsamples.  The solid line in each panel is the model (equation
\ref{wpmod}), fit to the data for that subsample; the fit parameters are given
in \tabref{Mdeptab}.  The dashed lines show the model from the brightest
subsample for each SED class.}

\fig{11}{rMdepfig}
{Scaled comoving correlation length $\rscl$ vs. median $\mnot$ for each
SED-$\mnot$ subsample.  Each subsample has $\zminm\le z<\zmaxm$.}

\fig{12}{rMdepfitfig}
{One- and $2\sigma$ ($\Delta\chi^2=$ 2.30 and 6.17) error contours for the fits
of equation (\ref{rmmod}) for the SED-$\mnot$ subsamples.}

\fig{13}{gamMdepfig}
{Correlation function slope $\gamma$ vs. median $\mnot$ for the SED-$\mnot$
subsamples.}

\fig{14}{gamMdepfitfig}
{One- and $2\sigma$ ($\Delta\chi^2=$ 2.30 and 6.17) error contours for the fits
of equation (\ref{gammmod}) for the SED-$\mnot$ subsamples.}

\end{document}